\documentclass[aps,prb,reprint,groupedaddress,showkeys]{revtex4-2}

\usepackage{graphicx}
\usepackage{threeparttable}
\usepackage{hyperref}

\begin{document}

\title{Spontaneous structural reconstructions and properties of ultrathin
triangular ZnSe nanoplatelets}

\author{Alexander I. Lebedev}
\email[]{swan@scon155.phys.msu.ru}
\affiliation{Physics Department, Lomonosov Moscow State University, 119991 Moscow, Russia}


\begin{abstract}

Two-dimensional (2D) materials have revolutionized all areas of development
of high-performance electronic devices. In particular, the unique electronic and
optical properties of II--VI semiconductor nanoplatelets have been found to be very
promising for optoelectronics. However, not all properties of this intriguing
class of materials are yet known.
A new, previously unknown hexagonal 2D structure of ZnSe nanoplatelets whose energy
is lower than the energies of all previously studied systems is found from
first-principles calculations. This structure appears as a result of spontaneous
reconstruction of the wurtzite structure and differs from it by the stacking order
of the bulk and near-surface Zn atomic layers. The phonon spectrum, electronic structure,
and band gap of the obtained nanoplatelets are calculated. The phonon spectra of
the nanoplatelets are in complete agreement with the spectra observed in experiment
and differ strongly from the vibrational spectra of ZnSe nanoclusters. The adsorption
of ZnCl$_2$ and $L$-cysteine molecules on the surface of the nanoplatelets is
studied and is shown to be accompanied by yet another spontaneous reconstruction
of the hexagonal structure into a tetragonal one and a new rearrangement of Zn
atoms in the near-surface layers. Calculations of the natural optical activity
of nanoplatelets covered with $L$-cysteine reveal an increase in the specific
(calculated per chiral molecule) optical activity, which is especially strong for
the Janus structures, as compared to the free $L$-cysteine molecule.

Published in \texttt{J. Phys.~Chem.~C 129, 7012 (2025), DOI: 10.1021/acs.jpcc.4c08561}

\end{abstract}

\maketitle

\section{Introduction}

Zinc selenide ZnSe is a typical representative of direct-gap II--VI semiconductors
possessing unique electronic and optical properties and being of great interest for
the development of optoelectronic devices such as light-emitting diodes, lasers, and
photoelectric converters of solar energy. Many papers have been devoted to studies
of low-dimensional structures made of zinc selenide. Quantum dots and nanowires were
studied in Refs.~\cite{JChemPhys.85.2237,JPhysChemB.102.3655,AdvMater.17.2471,ChemMater.17.1296,
JAmChemSoc.136.11121,MaterLett.159.229,JAmChemSoc.140.14627,CrystEngComm.21.2955,ACSNano.14.3847}.
First inorganic--organic hybrid nanostructures containing quasi-two-dimensional (2D) layers
of ZnSe with thicknesses of one monolayer (1ML) and 2ML were synthesized back in
2006~\cite{JAmChemSoc.129.3157}. Later, papers devoted to the experimental study of
colloidal ZnSe nanoplatelets appeared~\cite{NatureCommun.3.1057,
MaterLett.99.172,InorgChem.54.1165,JPhysChemLett.10.3465,ACSNano.14.3847,
Materials.16.1073,JPhysChemC.127.13112,JMolecLiquids.398.124187}.
In Ref.~\cite{NatureCommun.3.1057}, 0.91~nm thick (four atomic layers) ZnSe
nanoplatelets with a cubic zinc-blende structure with (110) orientation were
synthesized, and their activity with respect to photocatalytic water
splitting was demonstrated. Rectangular nanoplatelets of the wurtzite ZnSe
modification with a thickness of 1.4~nm and [$11{\bar 2}0$] orientation were
synthesized in Ref.~\cite{MaterLett.99.172}, and the conditions for the
formation of the nanoplatelets packets with a lamellar structure were established.
In Ref.~\cite{InorgChem.54.1165}, magic ZnSe nanoclusters, which are precursors
for formation of nanoplatelets, were synthesized and isolated, and their
transformation into the wurtzite ZnSe nanoplatelets was studied.
Rectangular ZnSe nanoplatelets of wurtzite modification with a thickness of 1.39~nm
and [$11{\bar 2}0$] orientation were synthesized in Ref.~\cite{JPhysChemLett.10.3465},
and the reason for the stability of nanostructures of exactly
this thickness (8~atomic layers) was explained based on first-principles calculations.
In Ref.~\cite{JMolecLiquids.398.124187}, the circular dichroism induced in wurtzite
ZnSe nanoplatelets by thermotropic liquid crystals, in which
the chirality of the cholesteric phase exists in a narrow temperature range, was
studied. A possible application of ZnSe nanoplatelets as cathodes for lithium
batteries was discussed in Ref.~\cite{Nanoscale.9.17303}.

\begin{figure}
\centering
\includegraphics[scale=0.9]{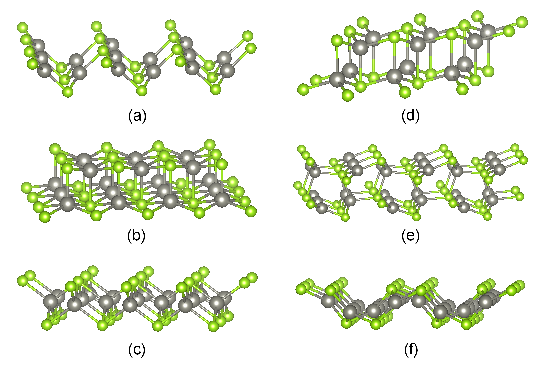}
\caption{\label{fig1}Side view of all 2ML ZnSe nanoplatelets studied in this work. (a) V-ZnSe,
(b) t-ZnSe, (c) t2-ZnSe, (d) tr-ZnSe, (e) WZ100-ZnSe, and (f) WZ110-ZnSe. It should be noted that
the WZ110 2ML structure spontaneously transforms to the t-ZnSe one when performing calculations
using PAW pseudopotentials. Se atoms are shown in green and Zn atoms are gray.}
\end{figure}

Theoretical studies of ZnSe nanoplatelets have shown the following. In
Ref.~\cite{JMaterChemA.2.17971}, the properties of three possible 2D structures
were described: of a flat hexagonal honeycomb structure 1ML thick ($P\bar{6}m2$
space group) which turned out to be dynamically unstable; the V-phase
(Fig.~\ref{fig1}(a))---a strongly corrugated orthorhombic structure with the
$Pmn2_1$ space group, into which the honeycomb structure transforms upon the phase
transition with a doubling of the unit cell volume; and a tetragonal t-phase
(Fig.~\ref{fig1}(b))---a 2ML thick nanoplatelet with
the $P4/nmm$ space group. The V- and t-phases of nanoplatelets turned out to be
dynamically stable, and the t-phase was the most energetically favorable. Both
of these phases were semiconductors whose band edge energy positions allowed
them to be used for photocatalytic water splitting. An evolutionary search for
configurations of possible 2D structures of ZnSe~\cite{AdvSci.2.1500290} has
found three stable structures: two already known phases (a hexagonal honeycomb
structure of 1ML thickness and the t-phase of ZnSe of 2ML thickness) and a new
pseudohexagonal phase ph-ZnSe of 2ML thickness with the $Pmn2_1$ space group
(we shall show later that ph-ZnSe and V-ZnSe are the same phase).
The t-phase had the lowest energy among the latter two. An interesting feature
of the ph-ZnSe phase was its tendency to transform into t-ZnSe at 300~K and
hydrostatic compression above 500~MPa, which was demonstrated using molecular
dynamics simulations. In Ref.~\cite{ACSApplMaterInterf.7.1458}, a new
tetragonal structure t2-ZnSe of 2ML thickness with the $P4/nmm$ space group
(Fig.~\ref{fig1}(c)) was found.%
    \footnote{It should be noted that there are \emph{two} different structures with
    the same $P4/nmm$ space group for 2ML thick nanoplatelets. The structure of
    the t-phase has one vertical Zn--Se bond and Se coordination number of 5,
    whereas in the new phase all interatomic bonds are oblique and Se coordination
    number is 4 (see Fig.~\ref{fig1}). To avoid further confusion, we shall call
    the second phase t2-ZnSe. The positions of atoms in these structures
    are given in Table~S1 of the Supporting Information.}
The effect of biaxial stress on the properties of 1ML thick nanoplatelets was investigated
in Ref.~\cite{JApplPhys.125.082540}. It should be noted that the results of phonon
spectra calculations for the hexagonal honeycomb structure of ZnSe contradict each other:
according to Ref.~\cite{JApplPhys.125.082540}, this structure is dynamically stable,
whereas its strong instability at the~$K$ and $M$ points of the Brillouin zone was found
in Ref.~\cite{PhysRevB.92.115307} (our calculations confirmed this conclusion).
Contradictions also concern the character of the band structure of honeycomb ZnSe:
according to Refs.~\cite{PhysLettA.381.663,PhysStatSolidiB.260.2300046}, it is
indirect-band, whereas other calculations~\cite{PhysRevB.92.115307,JApplPhys.125.082540}
indicate that it is direct-band. As expected, the strong quantum size effect is manifested
in all nanoplatelets.

The study of 1D structures (nanoribbons and nanotubes) that can be constructed
from 2D nanoplatelets was the subject of Ref.~\cite{PhysChemChemPhys.20.24453}.
Changes in the electronic, optical, and magnetic properties of t-ZnSe
nanoplatelets caused by doping with isoelectronic and non-isoelectronic impurities
were studied in Refs.~\cite{ChemPhysChem.17.1993,JAlloysComp.695.1392,MaterTodayChem.4.40}.
The effect of donor and acceptor impurities on the properties of ph-ZnSe was
investigated in Ref.~\cite{IntJHydrogenEnergy.46.34305}. The electronic structure
and optical properties of t-ZnS/ZnSe heterostructures were studied
in Ref.~\cite{PhysChemChemPhys.20.9950}.

The properties of two more 2ML-thick 2D ZnSe structures, WZ100 and WZ110
(Figs.~\ref{fig1}(e) and \ref{fig1}(f)),
which were cut from the wurtzite structure along the ($10{\bar 1}0$) and ($11{\bar 2}0$)
planes, were studied in Ref.~\cite{JMaterChemC.5.4505}. Both phases had an
orthorhombic structure. Among them, the ($11{\bar 2}0$) cut was energetically more stable.

It should be noted that the finding of a number of potential 2D structures whose
dynamic stability is confirmed by first-principles calculations indicates a broad
manifestation of metastability in these systems and the probability that the most
stable structure among them has not yet been found.

By lowering the growth temperature and changing the composition of the reaction mixture,
the authors of Ref.~\cite{Materials.16.1073} succeeded in obtaining a modification
of ZnSe in the form of thin triangular platelets with a thickness of only 0.6~nm and
lateral dimensions of about 200~nm. Their absorption spectra were characterized by
the appearance of two exciton absorption lines at 293 and 281~nm, which were attributed
to electronic transitions involving the sub-bands of light and heavy holes. According
to electron diffraction data, the hexagonal structure of the nanoplatelets was
identified as a wurtzite one. Its thickness, estimated by transmission electron
microscopy~\cite{Materials.16.1073} and atomic force microscopy~\cite{JPhysChemC.127.13112},
was 2.5ML (it was assumed that the nanoplatelets were covered with an additional
selenium layer to ensure symmetry). Thus, in this work, the 8-layer rule proposed
in Ref.~\cite{JPhysChemLett.10.3465} was experimentally refuted. Adsorption of
$N$-acetyl-$L$-cysteine on the surface of these nanoplatelets induced chirality in
them with the dissymmetry factor of 3.22$\times$10$^{-3}$. The phonon spectra of
the obtained ZnSe nanoplatelets were studied by infrared (IR) and Raman spectroscopy
in Ref.~\cite{JPhysChemC.127.13112}.

In connection with the above, the following should be noted. Magic-size nanoclusters
are believed to be the precursors of the nanoplatelets. The appearance of ZnSe nanoparticles
characterized by absorption bands near 280 and 291~nm was previously explained
by the formation of such nanoclusters~\cite{JPhysChemC.114.21921,InorgChem.54.1165,ACSNano.14.3847},
and the peaks at 280 and 291~nm were even attributed to different
nanoclusters~\cite{JPhysChemC.114.21921}. Structures with triangular geometry
formed by nanoclusters have already been observed in Ref.~\cite{ACSNano.14.3847}.
The temperature regimes of the synthesis of magic nanoclusters were close to
those used in Ref.~\cite{Materials.16.1073}. The following question arises: are the
objects obtained in Ref.~\cite{Materials.16.1073} self-assembled structures built
from nanoclusters or they are nanoplatelets? When studying the triangular
objects obtained in Ref.~\cite{Materials.16.1073}, two facts were noted that
testify in favor of nanoplatelets~\cite{Grafova_diss}. These were:
1)~the appearance in the X-ray diffraction pattern, in addition to the reflexes
indicating the formation of nanoparticles packets, of several additional fairly
narrow reflexes indicating the appearance of a long-range order in the arrangement
of atoms, and 2)~the fact that the stage of nanocluster formation, marked by
the appearance of optical absorption bands at 269 and 280~nm, was completed
during the synthesis process before the appearance of the phase under discussion.

In this work, a new, previously unknown 2D structure of ZnSe nanoplatelets whose
energy is lower than the energies of all previously studied systems is found from
first-principles calculations. The phonon spectra, electronic structure, and
band gaps of the nanoplatelets are calculated. The predicted phonon spectra of the
nanoplatelets agree well with the spectra observed in experiment and differ
strongly from the vibrational spectra of ZnSe nanoclusters. The adsorption of
ZnCl$_2$ and $L$-cysteine molecules on the surface of the nanoplatelets is studied,
and the natural optical activity of nanoplatelets covered with $L$-cysteine is
calculated. An increase in the specific (calculated per chiral molecule) optical
activity as compared to that of the free $L$-cysteine molecule is revealed. This
effect is especially strong for the Janus structures.

\section{Computational Details}

The electronic structure of bulk ZnSe and various nanostructures made from it
were calculated from first principles within the density functional theory using
the ABINIT software package~\cite{abinit3}. The local density approximation (LDA)
and PAW pseudopotentials taken from Ref.~\cite{ComputMaterSci.81.446} were used
in the calculations. The plane wave cutoff energy was 30~Ha (816~eV). The
nanoplatelets were modeled using supercells with periodic boundary conditions
in the $xy$ plane and a vacuum layer of $\geq$25~{\AA} separating the individual
nanoplatelets. The unit
cell parameters and atomic positions were relaxed until the forces acting on
the atoms became less than $5 \cdot 10^{-6}$~Ha/Bohr (0.25~meV/{\AA}) with an
accuracy of the total energy calculation better than that of 10$^{-10}$~Ha. To
reduce the systematic underestimation of the lattice parameter typical of the LDA
approximation and to solve a problem of the effective charge calculation in the
PAW approach, phonon spectra calculations were performed using the ONCVPSP
norm-conserving pseudopotentials~\cite{PhysRevB.88.085117} constructed for the
PBEsol functional~\cite{PhysRevLett.100.136406} with a cutoff energy of 49~Ha.

Since the spin-orbit interaction has virtually no effect on the geometry of the
structures, the calculations of the equilibrium geometry were performed without
taking spin into account. The electronic structure and spinor wave functions
were calculated with the spin-orbit interaction turned on.
Since ZnSe is a direct-gap semiconductor with the band extrema at the $\Gamma$~point
of the Brillouin zone, when calculating the matrix elements of optical transitions
we were interested in optical transitions exactly at this point.

\section{Results and discussion}

\subsection{Electronic structure of 2.5ML thick ZnSe nanoplatelets}

\begin{figure}
\centering
\includegraphics[scale=1.0]{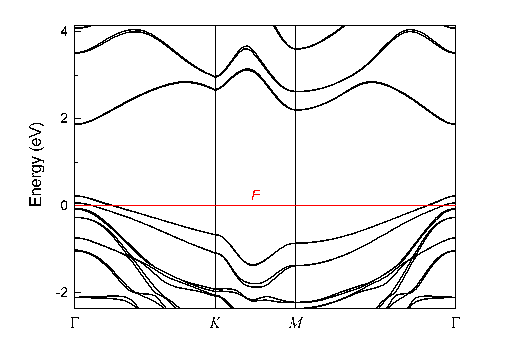}
\caption{\label{fig2}Band structure of a 2.5ML thick wurtzite ZnSe nanoplatelet
with the structure proposed in Ref.~\cite{Materials.16.1073}. The Fermi level~$F$
is located deep in the valence band.}
\end{figure}

The electronic band structure of a 2.5ML thick wurtzite ZnSe nanoplatelet with
the structure proposed in Ref.~\cite{Materials.16.1073} is shown in Fig.~\ref{fig2}.
It is seen that the character of the band structure of this nanoplatelet is
metallic because of the degeneracy of holes in the valence band. To test the
possibility of finding of a semiconductor phase among 2.5ML thick nanoplatelets,
all 12~possible configurations of filling the tetrahedral and octahedral voids
in the hexagonal closed-packed (HCP) and face-cubic centered (FCC) close-packed
structures formed by three layers of large Se atoms were analyzed (see details in
the Supporting Information). These configurations cover all possible structures
from the wurtzite one with the [0001] orientation to the zinc-blende and rocksalt
structures with the [111] orientation. Unfortunately, all configurations had a
metallic-type band structure. Since it is difficult to expect that such degenerate
semiconductor structures will exhibit the luminescence that was observed in the
experiment, one can conclude that the model of the nanoplatelet proposed in
Ref.~\cite{Materials.16.1073} is incorrect.

The appearance of degenerate holes in the electronic band structure of 2.5ML thick
nanoplatelets is easy to understand if we take into account that the addition to
the ZnSe nanoplatelet of one extra atomic layer of selenium, which has an incompletely
filled outer $p$ shell ([Ar]$3d^{10}4s^24p^4$) in the neutral state, will
leave two $p$ spin orbitals per primitive cell unoccupied and will result in the
appearance of the valence band states that are incompletely occupied with electrons.
In a certain sense, the observed situation resembles one that occurs in zinc-blende
CdSe nanoplatelets containing an additional atomic layer of Cd. In these
electrically neutral nanoplatelets, the $s$ electrons of Cd atoms pass to the
conduction band and result in a metallic-type electronic structure. To remove
these electrons from the conduction band, we needed two $X$ ligands that localize
these two electrons on themselves. The resulting redistribution of the electron
density in CdSe nanoplatelets can be seen from the behavior of the electrostatic
potential in them~\cite{JPhysChemC.127.9911}. In the case of an additional Se atomic
layer, we do not have enough electrons to completely fill the valence band.

Attempts to fill the vacant electron states by covering both sides of 2.5ML thick
wurtzite ZnSe nanoplatelets with hydrogen atoms were not very successful. After
adsorption, all nanostructures having the structures derived from the HCP one
retained a metallic band structure, which was due to the appearance in the forbidden
band of surface states (Figs.~S1 and S3 in the Supporting Information).
In the nanostructure with octahedral voids occupied like in the rocksalt structure,
the hydrogen adsorption transformed the hexagonal structure into a distorted one
with the $C2$ space group, in which the coordination numbers of atoms in the bulk
decreased from 6 to 4 and which became a semiconductor (Fig.~S2 in the Supporting
Information). The band structure of the nanoplatelet with the zinc-blende structure
(Fig.~S3 in the Supporting Information) was close to that obtained for the
wurtzite nanoplatelet. It is interesting that among the 2.5ML thick
hydrogen-passivated ZnSe nanoplatelets, the lowest-energy
structure was obtained for a HCP $AABAA$ structure that transformed from the
hexagonal structure into a slightly distorted (001)-oriented tetragonal one
upon the hydrogen adsorption; however, its symmetry was incompatible with the
triangular shape of the nanoplatelets. Passivation of the nanoplatelet surface
with Cl atoms resulted in the transformation of the unit cell into a highly
distorted structure resembling the
(001)-oriented zinc-blende one, which had an indirect-gap band structure.
Passivation of Zn-terminated 2.5ML wurtzite nanoplatelet by Cl atoms resulted
in the appearance of surface states in the forbidden gap~(Fig.~S4 in the
Supporting Information).

It is known that under normal conditions ZnSe has two stable modifications with
a cubic zinc-blende structure and a hexagonal wurtzite one. The zinc-blende
structure is a low-temperature modification, whereas the wurtzite structure
is a high-temperature one~\cite{JCrystGrowth.165.31} and is metastable
at room temperature. Under hydrostatic pressure of about 12~GPa, ZnSe
transforms into the rocksalt structure~\cite{JPhysChemSolids.56.521}. The
triangular shape of the nanoplatelets obtained in the experiment suggests that
their structure can be built from the structures of wurtzite (with the
orientation (0001)), zinc-blende, or rocksalt (with the orientation (111)).
In the above calculations of the nanoplatelets with a thickness of 2.5ML,
the objects whose structure was derived from the HCP one had the lowest
energy. The metallic character of the band structure of wurtzite nanoplatelets
with a half-integer number of monolayers forced us to consider ZnSe
nanoplatelets of stoichiometric composition.

\subsection{Structural transformation of wurtzite ZnSe(0001) nanoplatelets}

From the introduction it follows that wurtzite-based 2D structures with
the (0001) orientation have remained unexplored so far. Of two possible
variants of 2D nanostructures derived from the wurtzite one (Fig.~S5(a) in
the Supporting Information),
we will be interested in the energetically more favorable structure with the
coordination of surface atoms equal to 3 (Fig.~S5(c)). The structure
with a singly coordinated atoms on the surface (Fig.~S5(b)) has a
significantly higher energy and a metallic character of the band structure.

\begin{figure}
\centering
\includegraphics[scale=1.0]{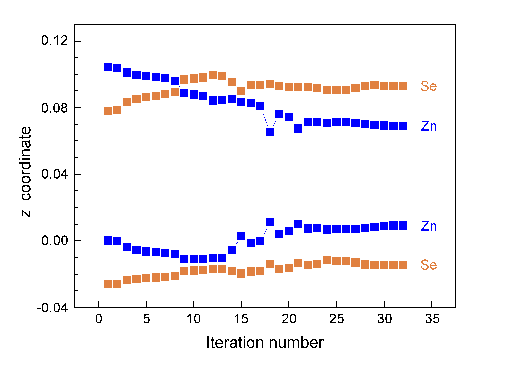}
\caption{\label{fig3}Dynamics of reconstruction of the initial wurtzite structure
into the tr-ZnSe one in a 2ML thick nanoplatelet.}
\end{figure}

It turned out that the wurtzite structure shown in Fig.~S5(c) is dynamically
unstable and undergoes an unusual structural rearrangement. Fig.~\ref{fig3}
shows the dynamics of the change in $z$ coordinates of atoms in a 2ML thick
nanostructure during the structural relaxation. It is seen that during the
relaxation, the Zn atoms located at one of the surfaces move into the bulk,
pushing the Se atoms onto the surface (Fig.~\ref{fig1}(d)). At the moment of
transition, the in-plane lattice parameter of the structure noticeably increases
and then relaxes to a new value that exceeds the initial value. Therefore, the
relaxed structure turns out to be terminated by selenium atoms, and its symmetry
is transformed from the polar $P3m1$ to the nonpolar $P{\bar 3}m1$. Note that
this reconstruction does not require thermal activation since its modeling
was performed with neglect of the thermal motion of the atoms.

We recalculated the structures of all previously published stable 2D ZnSe nanostructures
and added to them that of the new tr-ZnSe phase found in this work. The energies of
the obtained phases are given in Table~\ref{table1}, and the lattice parameters
and atomic coordinates for all structures are given in Table~S1 of the Supporting
Information. It turned out that the ph-ZnSe and V-ZnSe phases represent the same
structure. As follows from Table~\ref{table1}, the energy of the new tr-ZnSe structure
is lower than the energies of all of the previously studied nanostructures.

\begin{table}
\caption{\label{table1}Formation energies (per formula unit) of 2D ZnSe nanostructures
with a thickness of $N$~monolayers. $Z$ is the number of formula units in the unit
cell.}
\begin{threeparttable}[b]
\begin{ruledtabular}
\begin{tabular}{ccccc}
Phase & $N$ & $Z$ & Space group & Formation energy (eV) \\
\hline
t-    & 2   & 2  & $P4/nmm$ & $-$6.641 \\
V-    & 2   & 2  & $Pmn2_1$ & $-$6.778 \\
ph-   & 2   & 2  & $Pmn2_1$ & $-$6.778 \\
t2-   & 2   & 2  & $P4/nmm$ & $-$6.911 \\
WZ100 & 2   & 4  & $Pmc2_1$ & $-$6.996 \\
WZ110\tnote{1} & 4 & 8 & $Pca2_1$ & $-$7.034\tnote{1} \\
tr-   & 2   & 2  & $P{\bar 3}m1$ & {\bf $-$7.041} \\
\end{tabular}
\end{ruledtabular}
\begin{tablenotes}
\item[1]{\footnotesize When using the PAW pseudopotentials, the WZ110 phase with
$N = 2$, $Z = 4$ became unstable and relaxed to t-ZnSe. The origin why the energy
of the phase with $N = 4$ is lower than that of the t-ZnSe phase with $N = 2$
is the energy gain resulting from gluing of thinner nanoparticles.}
\end{tablenotes}
\end{threeparttable}
\end{table}

The obtained tr-ZnSe structure can be considered as a close-packed structure in which
\emph{all} tetrahedral voids (both of the $T_+$ and $T_-$ types) are filled with
Zn atoms. The energy gain occurs because the coordination numbers of
all atoms in this structure become equal to 4, although the coordination of the
surface Se atoms is strongly different from the tetrahedral one (see Fig.~\ref{fig1}(d)):
three Zn atoms are located at a distance of 2.46~{\AA} and one Zn atom at a distance
of 2.66~{\AA}. Note that the coordination of 4 for all atoms is also characteristic
of the t-ZnSe structure which until now was considered as the most energetically
favorable; in other structures, the coordination numbers of surface atoms are reduced
by~1. It is interesting that the discovered transformation of the wurtzite structure
into the tr-ZnSe one is of a fairly general character because similar transformations
were also established on the surface of (0001) wurtzite 2D structures with
a thickness of 2ML made of ZnS, ZnTe, CdS, CdSe, AlN, GaN, ZnO, BeO.

\begin{figure}
\centering
\includegraphics[scale=1.0]{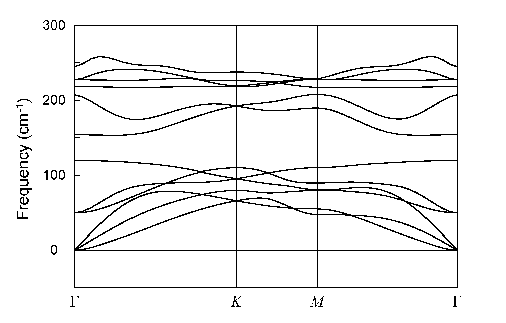}
\caption{\label{fig4}Phonon spectrum of a 2ML thick tr-ZnSe nanoplatelet.}
\end{figure}

Calculations of the phonon spectrum of the tr-ZnSe structure (Fig.~\ref{fig4})
confirm its dynamic stability.

A similar rearrangement of the surface structure was also observed in 3ML thick
ZnSe nanoplatelets. In 4ML thick ZnSe nanoplatelets the reconstruction was even
more interesting. It started from the surface Zn atoms and was accompanied by an
increase in the in-plane lattice parameter, which facilitated the subsequent
rearrangement of atoms \emph{in the bulk} of the nanoplatelet (Fig.~S6 in the
Supporting Information). Thus, the equilibrium configuration of a 4ML thick
nanoplatelet is a pair of 2ML thick nanoplatelets glued together. The cohesive
energy estimated from the difference in total energies of a 4ML thick nanoplatelet
and two isolated 2ML thick nanoplatelets is 126~meV per area of the unit cell. This
means that to prevent gluing, the surface of the nanoplatelets should be protected
by a ``coat'' of ligands. In thicker nanoplatelets (6ML), the large thickness
of the unreconstructed layer restricts the in-plane lattice parameter from
increasing, so the reconstruction occurs only on the surface and does not extend
into the depth of the nanoplatelet. Nevertheless, calculations show the possibility
of a further noticeable ($\sim$40~meV per formula unit) gain in energy of the
structure upon its complete reconstruction. One can suppose that upon thermal
activation, structural reconstruction may occur in the bulk of even thicker
nanoplatelets.

Although the above calculations have been performed for tr-ZnSe nanoplatelets
in the vacuum, additional calculations showed that their structure remains unchanged
in the presence of hexane molecules (nonpolar solvent).

\subsection{Electronic structure and optical transitions in tr-ZnSe nanoplatelets}

\begin{figure}
\centering
\includegraphics[scale=1.0]{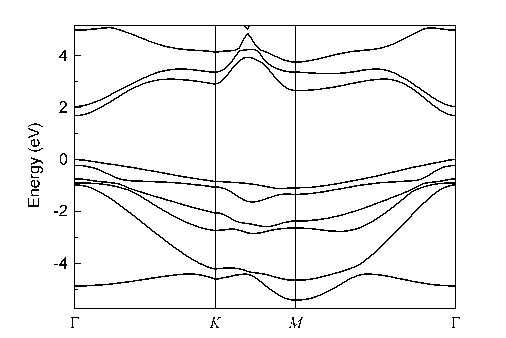}
\caption{\label{fig5}Band structure of a 2ML thick tr-ZnSe nanoplatelet.}
\end{figure}

\begin{table}
\caption{\label{table2}Squares of matrix elements of optical transitions averaged
over spin variables for a 2ML thick tr-ZnSe nanoplatelet.}
\begin{ruledtabular}
\begin{tabular}{ccc}
Transition & \multicolumn{2}{c}{Matrix elements (at. units)} \\
\cline{2-3}
           & $|P_x|^2 = |P_y|^2$ & $|P_z|^2$  \\
\hline
    HH1 ($E_u$)        $\to$ CB1 ($A_{2u}$) &  0    & 0 \\
    LH1 ($E_u+A_{2u}$) $\to$ CB1 ($A_{2u}$) &  0    & 0 \\
    HH2 ($E_g$)        $\to$ CB1 ($A_{2u}$) &  0.47 & $8 \cdot 10^{-5}$ \\
    SO1 ($A_{2u}$)     $\to$ CB1 ($A_{2u}$) &  0  & 0 \\
    LH2 ($E_g$)        $\to$ CB1 ($A_{2u}$) &  0.44 & $5 \cdot 10^{-5}$ \\
    HH1 ($E_u$)        $\to$ CB2 ($A_{1g}$) &  0.34 & $7 \cdot 10^{-5}$ \\
    LH1 ($E_u+A_{2u}$) $\to$ CB2 ($A_{1g}$) &  0.31 & $7 \cdot 10^{-3}$ \\
    HH2 ($E_g$)        $\to$ CB2 ($A_{1g}$) &  0    & 0 \\
    SO1 ($A_{2u}$)     $\to$ CB2 ($A_{1g}$) &  $1.6 \cdot 10^{-2}$ & 0.104 \\
    LH2 ($E_g$)        $\to$ CB2 ($A_{1g}$) &  0    & 0 \\
\end{tabular}
\end{ruledtabular}
\end{table}

Calculations of the band structure of ZnSe nanoplatelets of the new hexagonal
modification gave unexpected results. The band structure of 2ML-thick tr-ZnSe
nanoplatelets calculated taking into account the spin-orbit interaction is shown
in Fig.~\ref{fig5}. It is seen that the extrema of the conduction and valence
bands are located at the $\Gamma$~point, i.e. direct optical transitions are
possible in the semiconductor. However, an analysis of the projections of spinor
wave functions onto a set of basis functions of the ${\bar 3}m$ point group,
carried out using the technique~\cite{arXiv.2307.15534}, revealed
the following. The lowest conduction subband at the $\Gamma$~point has
the $A_{2u}$ symmetry, and the second conduction subband has the $A_{1g}$ one.
The upper subband of the valence band has the $E_u$ symmetry. The other subbands,
in descending order of energy, are: a mixture of $94\%E_u+6\%A_{2u}$, pure $E_g$,
a mixture of $86\%A_{2u}+14\%E_u$, almost pure $E_g$, and $A_{1g}$. From an
analysis of the dependence of the band energies on the adiabatic turning on
the spin-orbit interaction (Fig.~S7 in the Supporting Information) it follows
that the four upper valence subbands are pairs of light and heavy hole subbands
with size-quantization level numbers of 1 and~2. From the found symmetries of
bands it follows that the dipole-allowed optical transitions from two upper
valence subbands HH1 and LH1 are possible only to the \emph{second}
size-quantization level of the conduction band; transitions to the lowest-energy
conduction subband are forbidden because the bands have the same parity.
Calculations of the matrix elements of optical transitions (Table~\ref{table2})
confirm that this is
indeed the case. Optical transitions to the second subband of the conduction
band from the two upper valence subbands are strong in the $xy$ polarization
and weak in the $z$ polarization. Allowed transitions to the first conduction
subband appear only from the third valence subband and are strong in the $xy$
polarization.

The band structure and optical transitions for a 4ML thick tr-ZnSe nanoplatelet
have similar features (see Fig.~S8 and Table~S2 in the Supporting Information).

It is interesting that forbidden optical transitions between the nearest
subbands were typical not only for tr-ZnSe but also for t-ZnSe and t2-ZnSe
nanostructures (the symmetries of four upper subbands of the conduction band and
four lower subbands of the valence band for all nanostructures studied in this
work are given in Table~S3 of the Supporting Information). The calculations of
the optical matrix elements confirm this conclusion.

As the band gap is always systematically underestimated in the LDA approximation,
we calculated the band gap $E_g$ using the hybrid PBE0
functional~\cite{JChemPhys.105.9982}. The calculations were performed without
taking into account the spin-orbit interaction, and corrections for the valence
band edge shift due to the spin-orbit splitting were added from its separate
calculation (the HH--LH splitting was 0.395~eV for bulk ZnSe, the HH1--LH1
splitting was 0.224~eV for the 2ML tr-ZnSe nanoplatelet and 0.172~eV for the
4ML tr-ZnSe nanoplatelet). For bulk ZnSe, the calculated $E_g$ value was 3.000~eV,
which was by $\sim$0.3~eV higher than the
experimental value. For 2ML thick nanoplatelets $E_g$ was found to be equal to
3.785~eV, for 4ML thick nanoplatelets it was 3.687~eV. However, since the optical
transitions between the nearest valence and conduction subbands are forbidden
by selection rules, the calculated energy of the first allowed optical transition
in a 2ML thick tr-ZnSe nanoplatelet is 0.227~eV higher, i.e. equal to 4.012~eV.
For comparison, in the experiment the first absorption peak had an energy
of 4.23~eV, and the HH1--LH1 splitting was approximately 0.18~eV.

The discrepancy between the calculated and experimental $E_g$ data can be
explained by the self-interaction error characteristic of the density functional
theory~\cite{PhysRevB.23.5048}. It underestimates the energy gaps between
the electron levels and the vacuum level and may significantly delocalize the
wave function of the conduction band in the $z$~direction, thus resulting in
an underestimation of the calculated quantum size effect.

\subsection{Phonon spectrum of 2ML thick tr-ZnSe nanoplatelets}

It is known that zinc-blende ZnSe is characterized by two modes of optical
vibrations: 205~cm$^{-1}$ (TO phonon) and 252~cm$^{-1}$ (LO phonon)~\cite{Adachi2009}.
Calculations of the phonon spectrum of zinc-blende ZnSe using the norm-conserving
RRKJ pseudopotentials~\cite{PhysRevB.41.1227} and the LDA approximation yielded
frequencies of 216.4~cm$^{-1}$ (TO) and 254.1~cm$^{-1}$ (LO). The calculations
using PAW pseudopotentials yielded even higher frequencies. When using the ONCVPSP
norm-conserving pseudopotentials~\cite{PhysRevB.88.085117} along with the PBEsol
exchange-correlation functional, the frequencies
became equal to 209.4 cm$^{-1}$ (TO) and 246.3 cm$^{-1}$ (LO). Since these
pseudopotentials gave the best agreement of the calculated phonon frequencies
with the experiment, we will use the results obtained using the ONCVPSP+PBEsol
calculation scheme in what follows.

Calculations of the phonon frequencies and their effective charges for a 2ML
thick tr-ZnSe nanoplatelet reveal in the frequency range covered by the
experiment~\cite{JPhysChemC.127.13112} a strong TO phonon line at 207.8~cm$^{-1}$
($E_u$ symmetry), a weak TO phonon line at 212.1~cm$^{-1}$ ($A_{2u}$), a weak
LO phonon line at 213.0 cm$^{-1}$ (LO$_z$), and a strong LO phonon line at
234.5 cm$^{-1}$ (LO$_{xy}$). Two lines at 210 and 238~cm$^{-1}$ were observed
in the infrared spectra~\cite{JPhysChemC.127.13112}. Calculations of the
frequencies and intensities of the Raman lines find only one line at
216.3 cm$^{-1}$ ($E_g$ symmetry) in the measurement range, whereas the strongest
Raman line at 158.8 cm$^{-1}$ is outside it. In the experiment, the Raman line
was observed at 211 cm$^{-1}$~\cite{JPhysChemC.127.13112}.
A small discrepancy between the observed and calculated phonon frequencies
can be explained by a strain of thin nanoplatelets on the substrates:
calculations show that the in-plane biaxial stretching of the nanoplatelets
reduces the frequencies of all modes, and a stretching of about 1\% is enough
to obtain the agreement between the calculations and experiment. The presence
of the 211~cm$^{-1}$ line and the absence of the 238~cm$^{-1}$ line in the
Raman spectra are the arguments in favor of the centrosymmetry of the nanoplatelets.

Returning to the question raised in the introduction about the identification
of the observed triangular nanoobjects, it should be noted that
according to Ref.~\cite{InorgChem.58.1815}, the optical absorption bands
at 292 and 281~nm, close to those observed in Ref.~\cite{Materials.16.1073},
were attributed to Zn$_{13}$Se$_{13}$ nanoclusters. Our calculations of the
vibrational spectra of several magic nanoclusters showed that the frequencies
of their IR active vibrations are shifted to higher frequencies as compared to
nanoplatelets, reaching 292 cm$^{-1}$ (for the Zn$_6$Se$_6$ cluster) and even
360 cm$^{-1}$ (for the Zn$_{13}$Se$_{13}$ cluster with a cage-case structure).
Therefore, the good agreement between the frequencies of TO and LO modes in
the IR and Raman spectra and the calculated frequencies for
tr-ZnSe nanoplatelets confirms that the objects under discussion are nanoplatelets
and not self-assembled structures formed from magic nanoclusters.

\subsection{Adsorption of ZnCl$_2$ and $L$-cysteine on the surface of tr-ZnSe nanoplatelets}

As was shown in the experiment~\cite{Materials.16.1073,Grafova_diss}, the
triangular ZnSe nanoplatelets can adsorb both ZnCl$_2$ molecules and
$N$-acetyl-$L$-cysteine molecules on the surface. In the nanoplatelet model
proposed in Ref.~\cite{Materials.16.1073}, the adsorption of acetylcysteine
is problematic since for its effective binding, the surface must be covered
with metal atoms. The binding of both molecules could be possible on a nonpolar
(covered with both Zn and Se) wurtzite surface with the ($11{\bar 2}0$)
orientation, but these nanoplatelets have an orthorhombic structure, and it is
not clear why the nanoplatelets have the triangular shape. We shall show that
the structure of tr-ZnSe nanoplatelets found in the present work enables to
explain their adsorption properties observed in the experiment. This opportunity
arises due to the possibility of a significant reconstruction of the nanoplatelet
surface structure upon adsorption.

For our analysis, we chose ZnCl$_2$ and a simpler molecule, $L$-cysteine, because
the latter has the
same thiol anchor group as acetylcysteine. It turned out that neutral molecules
of both cysteine and ZnCl$_2$ are able to attach to tr-ZnSe nanoplatelets because
the \emph{charged} cysteine radicals and Cl$^-$ ions induce a new rearrangement
of the surface structure, in which the Zn atoms come to the surface to form
chemical bonds with sulfur or chlorine atoms (Fig.~S10 in the Supporting Information).
For cysteine, the lowest-energy configuration is one in which cysteine is ionized
at the S--H bond, and a hydrogen ion is inserted into the nanoplatelet at the
position previously occupied by the Zn atom; the configuration with a hydrogen
ion on the other side of the nanoplatelet has a slightly higher energy. The
adsorption energy calculated as the difference between the total energies of
the nanoplatelet with the adsorbed molecule and the sum of the total energies
of the free nanoplatelet and the cysteine molecule is 1.54~eV. The energy gain
upon covering the nanoplatelet on both sides with Cl$^-$ and (ZnCl)$^+$ ions
is 2.19~eV.

However, this is not all. It turned out that when any charged ion (both Cl$^-$
and cysteine) approaches one surface of the nanoplatelet, the angle between
the~$a$ and $b$~axes of its primitive cell changes, and from the hexagonal structure
it spontaneously and irreversibly transforms into the \emph{tetragonal} one with
the (001) orientation and atomic arrangement like in the zinc-blende structure
(Fig.~S11 in the Supporting Information). We have already observed something
similar above when having studied the adsorption of hydrogen and chlorine
atoms on the 2.5ML thick ZnSe nanoplatelets.
This seems very surprising since a structure with a third-order axis transforms
into a structure with a fourth-order axis. Apparently, this is possible due to
the extremely small thickness of the nanostructure. The ZnCl$_2$-covered
nanostructures are direct-band semiconductors with allowed optical transitions
and an LDA band gap of 2.06~eV. For cysteine-covered nanostructures, the valence
band extremum moves to the $X$~point because of the appearance of an extra band
produced by cysteine in the forbidden gap. It should be noted that the
possibility of transformation of the pseudohexagonal ph-ZnSe into the tetragonal
t-ZnSe has already been demonstrated earlier in the study of ZnSe nanoplatelets
using the molecular dynamics~\cite{AdvSci.2.1500290}, but in the cited paper it
occurred only under hydrostatic compression.

While the nanoplatelets with adsorbed ZnCl$_2$ retain their nonpolar symmetry
$P{\bar 4}m2$, the nanoplatelet with cysteine adsorbed at one side turns out
to be the Janus one. The properties of Janus CdSe nanoplatelets were previously
studied in Ref.~\cite{JPhysChemC.127.9911}. When cysteine is adsorbed on
both sides of the nanoplatelet, both Zn atoms come to the surface thus
forming bonds with sulfur, and the structure becomes highly distorted with an
angle between the $a$ and $b$ axes of 111.7$^\circ$, but remains stable.

\begin{table}
\caption{\label{table3}Specific natural optical activity of $L$-cysteine and
2ML thick tr-ZnSe nanoplatelets covered with $L$-cysteine.}
\begin{ruledtabular}
\begin{tabular}{cc}
Object & ${\rm Tr}~\Omega g_{\alpha\alpha}/3$ (Bohr$^4$) \\
\hline
Free cysteine molecule & +13.3 \\
One side covered nanoplatelet & $-$146.1 \\
Two side covered nanoplatelet & $-$12.7 \\
\end{tabular}
\end{ruledtabular}
\end{table}

Calculations of the natural optical activity (NOA) carried out using the
approach~\cite{PhysRevLett.131.086902} implemented in the ABINIT 10.0.3 program
showed that the specific optical activity (activity per chiral molecule) of 2ML
thick tr-ZnSe nanoplatelets covered on one side with $L$-cysteine is 11~times
higher than the specific activity of a free $L$-cysteine molecule (the
orientation-averaged (effective) gyration pseudotensor, ${\rm Tr}~\Omega g_{\alpha\alpha}/3$,
is given in Table~\ref{table3}). This means that the induced chirality in nanoplatelets
may be much stronger than the chirality of a molecule itself. The sign of the NOA in
this Janus nanoplatelet is negative. For a 2ML thick tr-ZnSe nanoplatelet covered with
$L$-cysteine on both sides, the NOA is much smaller and also negative. At the same
time, the
specific NOA for the ($11{\bar 2}0$)-oriented ZnSe nanoplatelet covered with one molecule
of cysteine is positive (+119.4~Bohr$^4$). The detected difference in signs of the
optical activity correlates with the difference in signs of the circular dichroism
found upon the adsorption of cysteine on CdSe nanoplatelets with zinc-blende and
wurtzite structures~\cite{NanoLett.18.6665}. For zinc-blende CdSe nanoplatelets, the
calculated specific NOA induced by the adsorption of $L$-cysteine on two (001) surfaces
of a 2ML thick nanoplatelet is positive (+286.8~Bohr$^4$), and the specific NOA for
the corresponding Janus nanoplatelet (covered on one side) is 30\% stronger and also
positive (+370.2~Bohr$^4$). The effect of a significant increase in the specific
optical activity of chiral molecules adsorbed on Janus nanoplatelets deserves further
research.

\subsection{Conclusions}

In this work, a new, previously unknown 2D structure of ZnSe nanoplatelets whose
energy is lower than the energies of all previously studied systems has been found
from first-principles calculations. This structure appears as a result of
spontaneous reconstruction of the wurtzite structure and differs from it by the
stacking order of bulk and near-surface Zn atomic layers. The phonon spectrum,
electronic structure, and band gap of the nanoplatelets are calculated.
The phonon spectra of the nanoplatelets are found to be in complete agreement
with the spectra observed in the experiment and differ strongly from the vibrational
spectra of nanoclusters. It was shown that the adsorption of ZnCl$_2$ and cysteine
molecules on the surface of the nanoplatelets is accompanied by yet another
reconstruction of the hexagonal structure into a tetragonal one and a new
rearrangement of Zn atoms in the near-surface layers. Calculations of the natural
optical activity of ZnSe nanoplatelets covered with $L$-cysteine reveal a significant
increase in the specific (per chiral molecule) optical activity as compared to
free-standing $L$-cysteine molecules.

\section*{Conflict of interest}

The author declares that he has no conflict of interest.

\bigskip

\section*{Supporting information}

The supporting information is available at
\url{https://pubs.acs.org/doi/10.1021/acs.jpcc.4c08561}.


\begin{acknowledgments}
The work was financially supported by the Russian Science Foundation, grant~22-13-00101.
The author thanks R. B. Vasiliev for numerous fruitful discussions of various aspects
of the problem and for providing details of the triangular ZnSe nanoplatelets synthesis.
\end{acknowledgments}

\providecommand{\BIBYu}{Yu}

\end{document}